\documentclass[preprint,aps,showpacs]{revtex4}
\usepackage{graphicx} 
\usepackage{dcolumn}          
\usepackage{bm}               

\begin{document}
\preprint{APS/123-QED}

\title{Simultaneous surface acoustic wave and surface plasmon resonance
measurements: electrodeposition and biological interactions monitoring}

\author{J.-M Friedt}%
\email{friedtj@imec.be}
\affiliation{IMEC, MCP/BIO, Kapeldreef 75, 3001 Leuven, Belgium}
\author{L. Francis}%
\affiliation{IMEC, MCP/BIO, Kapeldreef 75, 3001 Leuven, Belgium}
\author{G. Reekmans}
\affiliation{IMEC, MCP/BIO, Kapeldreef 75, 3001 Leuven, Belgium}
\author{R. De Palma}
\affiliation{IMEC, MCP/BIO, Kapeldreef 75, 3001 Leuven, Belgium}
\author{A. Campitelli}
\email{campi@imec.be}
\affiliation{IMEC, MCP/BIO, Kapeldreef 75, 3001 Leuven, Belgium}
\author{U.B. Sleytr}
\affiliation{Center for Ultrastructure Research and Ludwig Bolzmann 
Institute for Molecular Nanotechnology, Universit\"at f\"ur 
Bodenkultur Wien, Gregor-Mendel-Str. 33, A-1180 Vienna, Austria}

\date{\today}

\begin{abstract}
We present results from an instrument combining surface acoustic wave (SAW) 
propagation and surface plasmon resonance (SPR) measurements. The
objective is to use two independent methods, the former based on adsorbed
mass change measurements and the latter on surface dielectric properties
variations, to identify physical properties of protein layers, and more 
specifically their water content. We display mass sensitivity calibration 
curves using electrodeposition of copper leading to a sensitivity in liquid
of 150$\pm15$~$cm^2/g$ for the Love mode device used here, and the application 
to monitoring biological processes. The extraction of protein layer 
thickness and protein to water content ratio is also presented for 
S-layer proteins under investigation. We obtain respectively 
4.7$\pm$0.7~nm and 75$\pm$15\%.
\end{abstract}



\maketitle

\section{Introduction}

Surface plasmon resonance (SPR) is a well accepted direct detection technique
for monitoring biological processes \cite{liedberg, gizeli1}. 
While ellipsometry is another well known method for analyzing thin film 
properties \cite{kasemo}, its use in liquid medium
for monitoring biochemical reactions is made difficult by the varying 
environment through which the probing light beam has to propagate. In the 
Kretschmann configuration, the laser generating the SPR evanescent wave is 
only propagating through the substrate, leading to a better control over
the influence of the various buffer solutions used during a protein
adsorption experiment. 

The use of various acoustic wave devices for monitoring bound mass changes
in liquid media is well known \cite{gizeli2}. Love mode devices,
based on the propagation of a guided shear acoustic wave, present
sensitivity improvements over the more usual quartz crystal microbalance 
\cite{gizeli3} as well as a compatibility with measurements in liquids.

We take advantage of the unique geometrical setup of the surface acoustic
wave (SAW) device which
leaves the backside of the quartz wafer free of electrodes to inject
a laser in order to generate an evanescent surface plasmon on the 
gold coated sensing area. Such a setup enables
simultaneous estimates of the bound mass and dielectric surface properties
changes during electrochemical and biochemical reactions occurring on the 
sensing electrode \cite{bailey,johan1}.

Estimates of water content in protein layers is an important topic in
the development of biosensors, since the detection sensitivity is
directly related to the number of active sites on the surface of the
sensor to which the receptor molecules are bound. When physical
measurements of this bound layer are performed, such as via SPR or
acoustic measurements, a large water content will lead to overestimates
of the potential detection limit of the biosensor due to an overestimate
of number of molecules bound to the surface. From a more fundamental
point of view, determining water level content should provide more accurate
physical parameters of the protein layer itself such as density and
optical index by allowing the subtraction of the influence of water once
it has been identified.

\section{Experimental data}

We use a modified commercial SPR instrument (IBIS II, IBIS Technologies BV,
The Netherlands)
to detect the surface plasmon resonance angle after replacing the
gold coated glass slide by a Love mode SAW device (Fig. \ref{fig1}). 
The excitation laser in
this instrument has a wavelength of 670~nm. All reflected intensity {\em vs.} 
angle curves (6~$^o$ angle span recorded on a 200~pixels CCD array) were recorded 
and later fitted by a polynomial to extract with high accuracy the position 
of the dip. The acoustic wave
device is made of a 500~$\mu$m thick ST-cut quartz wafer on which 200~nm
thick sputtered $Al$ interdigitated electrodes are patterned. The surface is 
coated by a
1.13~$\mu$m PECVD silicon dioxide layer acting as a guiding layer, and
the 4.9$\times$5.4~mm$^2$ sensing area is coated with 10~nm $Ti$ and 50~nm
$Au$. This area acts both as a working electrode for electrochemistry or 
a grounded
electrode during biochemical experiments, as well as a supporting layer
for the surface plasmon resonance generation. The influence of the SAW device
substrate over the detection of the SPR is limited to interference patterns
due to the birefringence of the quartz, and the optical index mismatch between
the quartz and the deposited silicon dioxide layer. The former effect is 
reduced by orienting the optical axis of quartz so that it is in the
plane defined by the normal to the sensing surface and the wavevector of the
laser, thus minimizing the optical index difference between the ordinary
and the extraordinary axis. The remaining interference patterns lead
to fringes with a low enough contrast that the surface plasmon peak is 
easily identified and tracked during the experiments.

The phase of the SAW delay line is monitored at a fixed frequency using a 
network 
analyzer HP4396A at 123.200~MHz. The phase shift is converted to a frequency
shift thanks to the linear phase to frequency relationship recorded in the
Bode plot. The observed phase shift leads to a frequency shift (as would be
observed in a phase locked loop configuration) which in turn can be 
converted to an adsorbed mass change through the sensitivity of the
device.

We first electrodeposited copper on the surface in order to calibrate
the mass sensitivity of the acoustic wave device in liquid medium (Fig.
\ref{fig2}) \cite{me}. At
the same time, the SPR displayed resonance angle shifts due to
the varying potential \cite{kolb}. When the voltage applied by the 
potentiostat is above 0.2~V with respect to the $Cu$ pseudo reference
electrode, the SPR angle slightly shifts due to the electroreflectance
effect described by K{\"o}tz {\it et al.} \cite{kolb}. Below 0.2~V, under
potential deposition starts depositing a mono layer of $Cu$ atoms on 
the $Au$ surface as visible both in the phase shift of the SAW device
and as a reversal of the trend of the angle shift of the SPR. Below 0~V, a
rough $Cu$ layer is deposited, leading to a loss of the SPR peak and a large
phase shift of the SAW device due to the large added mass.
The rough, discontinuous film of copper clusters a few tens to hundreds
of nanometers high was observed by in-situ AFM imaging (data not shown
\cite{me}). Simulating the
influence of sub-wavelength metallic structures on the surface plasmon
resonance is a difficult task \cite{ebbesen,JACS} which we have 
not attempted to solve.
Simultaneously, the SAW device displays a frequency drop $\Delta f$ due to 
the added mass $\Delta m$ to the sensing electrode, leading with the 
combination of the current measurement from the potentiostat to a direct 
estimate of the mass sensitivity $S=\frac{\Delta f}{f_0}\frac{A}{\Delta m}$ 
where $A=0.49\times 0.54$~cm$^2$ is the area of the sensing working electrode
and $f_0=123.200$~MHz the base frequency of the SAW device. The electrodeposited
mass is deduced from the current measurement $I$ with a sampling period $\delta 
t$: $\Delta m=\frac{\sum_j I_j\times\delta t}{F}\times\frac{M}{n_e}$ where
$F=96485$~C is the Faraday constant, $M$ the molar weight of the
electrodeposited metal and $n_e$ the number of electrons exchanged for each 
metallic atom reduced. The resulting sensitivity for the Love mode device 
under consideration here is 150$\pm$15~cm$^2$/g.

Once the mass sensitivity of the SAW is calibrated, we adsorbed on the
same device a crystalline monolayer of S-layer proteins 
(100~$\mu$g/ml of the protein
SbpA of {\em Bacillus sphaericus} CCM 2177 in 0.5~mM tris, 10~mM $CaCl_2$,
pH=9 buffer) \cite{review,erika} while monitoring
simultaneously with the SAW device the mass variation and with SPR the
surface dielectric change. The adsorption and desorption steps, the
latter being possible with the use of 2~\% $NaOCl$, was repeated several
times (Fig. \ref{fig3}). A property of this protein which makes it suitable
for calibration of a new instrument is that it only forms a monolayer
and will not stack to multiple layers even at high concentrations.

\section{Discussion}

The parameters required for analyzing the SAW data are the layer thickness
$d$ and the layer density $\rho_{layer}$. The parameters required for 
analyzing the SPR data are the common layer thickness $d$ and the optical 
index $n$ of the adsorbed
layer. We thus have three free parameters and only two
measurements, namely the adsorbed mass per unit area and an SPR resonance
angle shift which is a function of both $n$ and $d$. The two options
are to add one more measurement parameter, such as
an SPR dip position as a function of wavelength which would lead to a
unique identification of $d$ and $n$ \cite{multil}, or to reduce the 
number of variables
by adding the assumption that the layer is made of a homogeneous mixture of
a proportion $x$ of protein and $1-x$ of water. The layer density
$\rho_{layer}$ is then $\rho_{layer}=x\rho_{protein}+(1-x)\rho_{water}$
where $\rho_{protein}$ is assumed to be equal in the $1.2$~g/cm$^3$ 
to 1.4~g/cm$^3$ range \cite{caruso,kasemo,hook2,marsch} and
$\rho_{water}=1$~g/cm$^3$ is the density of water. Similarly, we use then
the optical index of the layer $n_{layer}=xn_{protein}+(1-x)n_{water}$ with
the optical index of the protein layer assumed to be in the
$n_{protein}=1.45$ to $1.465$ range \cite{hook2,marsch,liedberg} and
that of water $n_{water}=1.33$. The remaining unknown parameters are then
the layer thickness $d$ and the water content proportion $1-x$. By
simulating a stack of planar multilayers (glass $n_{glass}=1.518$, 1200~nm 
silicon dioxide $n_{SiO_2}=1.45$, 2~nm titanium $n_{Ti}=2.76+i3.84$
\cite{palik}, 50~nm
gold $n_{Au}=0.14+i3.697$ \cite{palik}, proteins, water) for the angle shift as a function
of water content and layer thickness, one obtains a set of pairs of
values for these two parameters compatible with the observed angle shift
(Fig. \ref{fig4}, bottom).
Then, by calculating the mass per unit area $\Delta m=\rho d$, a comparison
with the experimentally observed adsorbed mass as seen from the phase
shift of the SAW device leads to a unique set of parameters both compatible
with the optical index change and the adsorbed mass (Fig. \ref{fig4}, top). 
In the case under
consideration here, an observed angle shift of 380 to 400~~m$^o$ and an 
adsorbed mass of 540 to 580~ng/cm$^2$ is only compatible with $x=0.75\pm0.15$ 
and $d=4.7\pm0.7$~nm. The thickness result is compatible with atomic
force microscope (AFM) measurements in liquid \cite{erika} while the mass 
per unit area is compatible with morphological data obtained by electronic 
microscopy \cite{review}. While the main source of 
uncertainty is due to the wide possible values of the density and optical 
index reported in the literature, the experimental results display good 
reproducibility. The water content is lower than that observed
by other authors for different kinds of proteins \cite{kasemo,marsch} but
compatible with the 30 to 78\% value cited in Ref. \cite{stenberg}. The
dense packing of protein is explained by the regular arrangement of
identical S-layer subunits in the p4 lattice \cite{weygand1,weygand2}.

\section{Conclusion}

We have shown here how the combination of Love mode SAW device with
SPR provides advantageous combinations of information on the bound mass
as well as on the dielectric changes when monitoring protein adsorption.
The resulting protein layer thickness and protein content percentage,
respectively $d=4.7\pm0.7$~nm and $x=75\pm 15$~\%, is in agreement with
independent AFM estimates. 

\section{Acknowledgments}

The SAW transducers were fabricated within the framework of a Belgian PhD
scholarship program (Fonds pour la formation \`a la Recherche dans l'Industrie 
et dans l'Agriculture -- FRIA). We wish to thank R. Giust (LOPMD, 
Besan\c con, France) for kindly providing SPR simulation routines.

\newpage

\begin{figure}[h!tb]
\caption{Experimental setup including, from bottom to top: the cylindrical
glass prism for injecting the laser, the quartz wafer patterned with
interdigital transducers for generating the acoustic waves and coated with a
thin (1.2~$\mu$m) $SiO_2$ layer, and the SU8 walls supporting the glass slides
capping preventing the contact of the liquid over the transducers area. We
indicate for clarity on the top graph the equivalent frequency shift (in kHz) and
on the bottom graph the equivalent mass and calculated sensitivity.}
\label{fig1}
\end{figure}

\begin{figure}[h!tb]
\caption{SAW sensitivity calibration curves using copper electrodeposition.
From bottom to top: potentiostat current and potential {\em vs.} a copper
wire acting as a pseudo reference electrode with the $Cu^{2+}$ ions in
solution~; SPR angle shift (channel 1 and 2 are separated by about 2~mm)~; and
surface acoustic wave phase monitored at a fixed frequency (123.200~MHz) and
converted to a frequency shift thanks to the linear phase to frequency
relationship (data not shown).}
\label{fig2}
\end{figure}

\begin{figure}[h!tb]
\caption{Simultaneous measurement of the SAW phase at 123.200~MHz and the SPR
angle shift during similar experiments: injection of buffer solution, 
injection of S-layer protein (100~$\mu$g/ml), injection of buffer and removal
of the proteins by injecting 2\% $NaOCl$.}
\label{fig3}
\end{figure}

\begin{figure}[h!tb]
\caption{Left: simulation of the SAW frequency shift for varying protein layer
thickness (abscissa) and proportion of proteins in the layer (from $x=10$\%
to $x=100$\% from right to left), and varying the density of the protein
layer within the values found in the literature (from 1.2~g/cm$^3$ as solid 
lines to 1.4~g/cm$^3$ as dots). Right: simulation of the SPR angle shift for
varying protein layer thickness (abscissa) and proportion of proteins in the
layer (from $x=10$\% to $x=100$\% from bottom to top), and varying the optical
index of the protein layer within the values found in the literature (from
1.45 as solid lines to 1.465 as dots). The mass density of the points lying in 
the region of experimentally observed angle shift have been indicated for added 
clarity.}
\label{fig4}
\end{figure}

\newpage
\includegraphics[width=15cm]{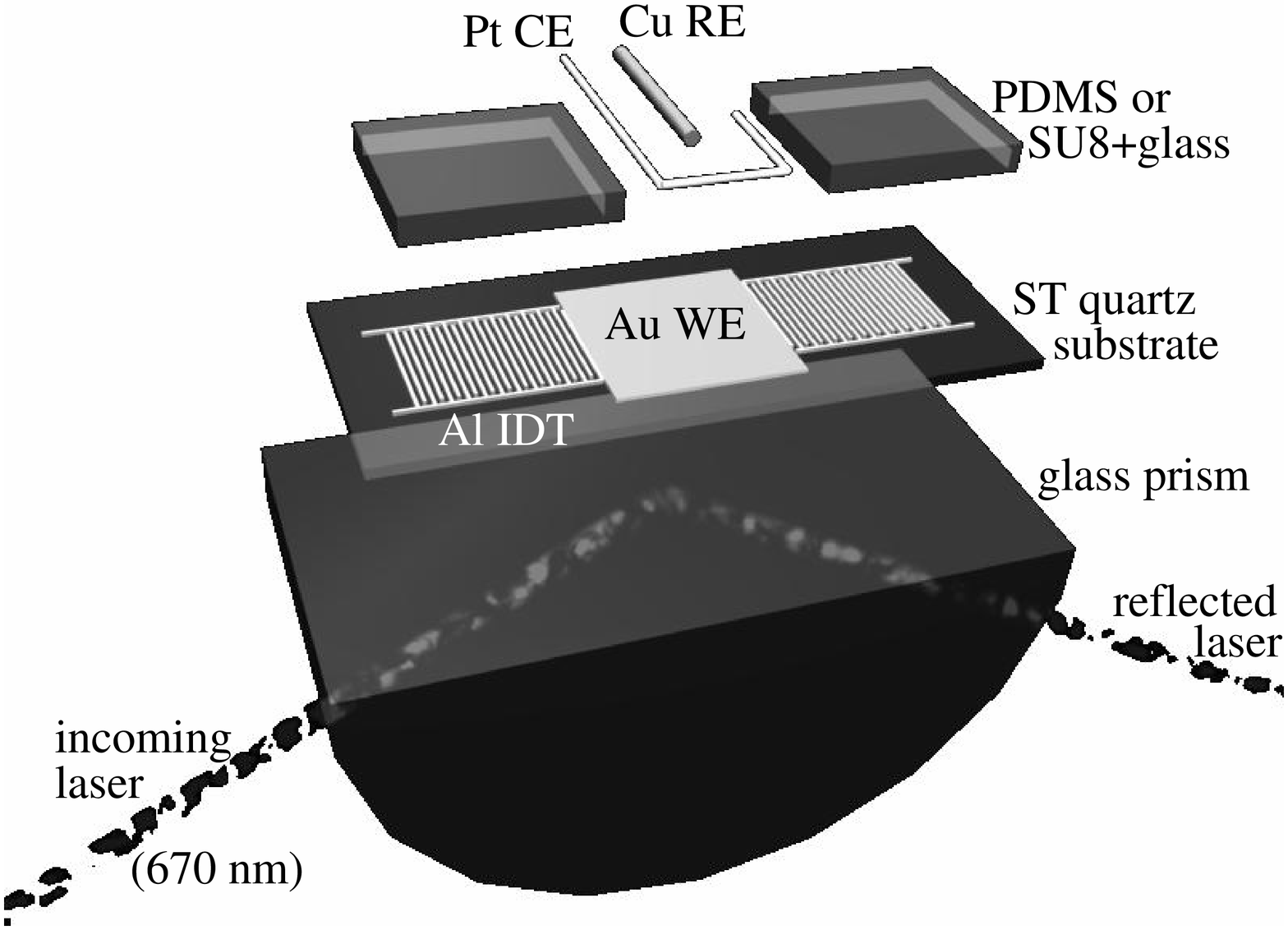}\newpage
\includegraphics[width=15cm]{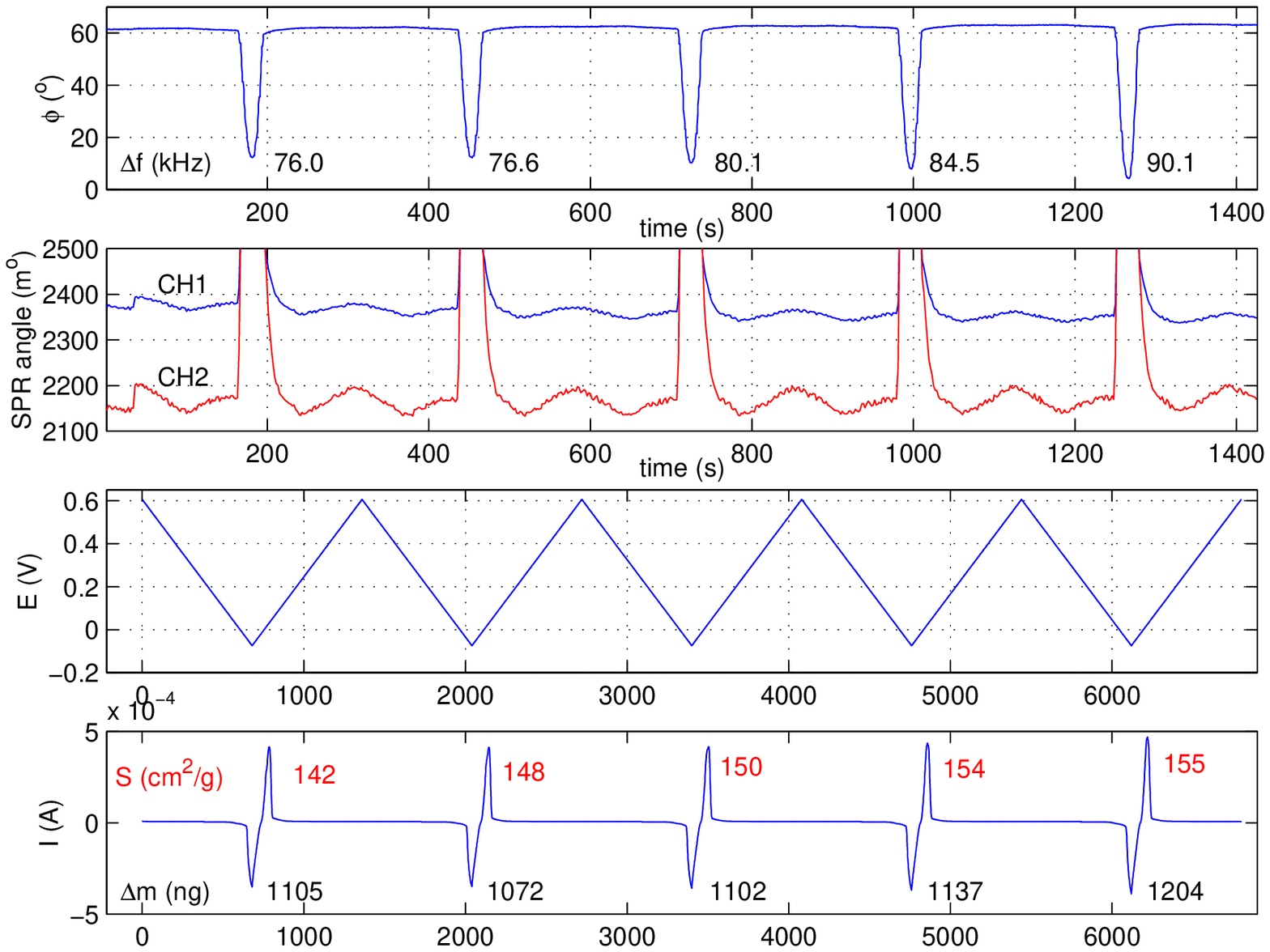}\newpage        
\includegraphics[width=15cm]{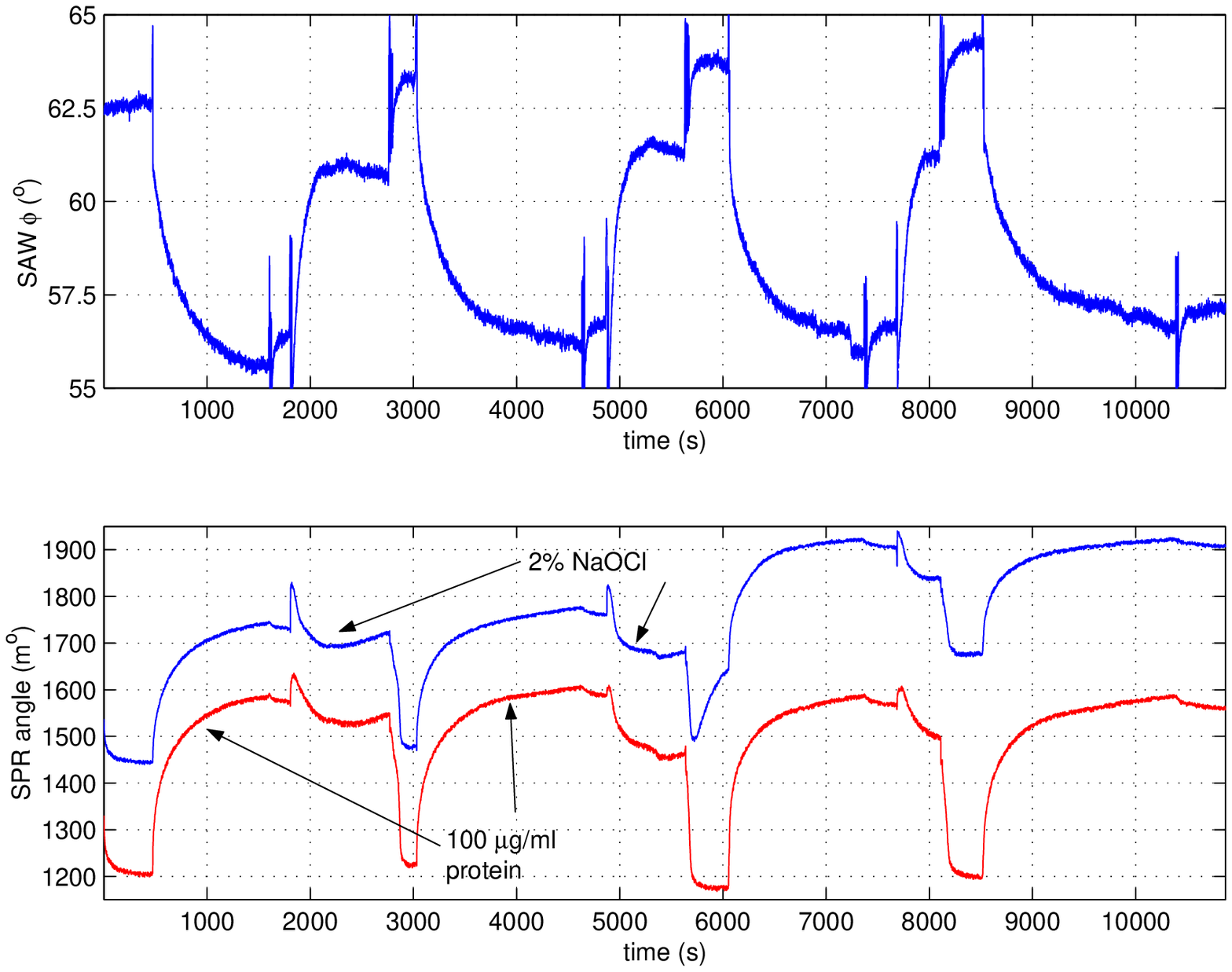}\newpage   
\noindent\includegraphics[width=15cm]{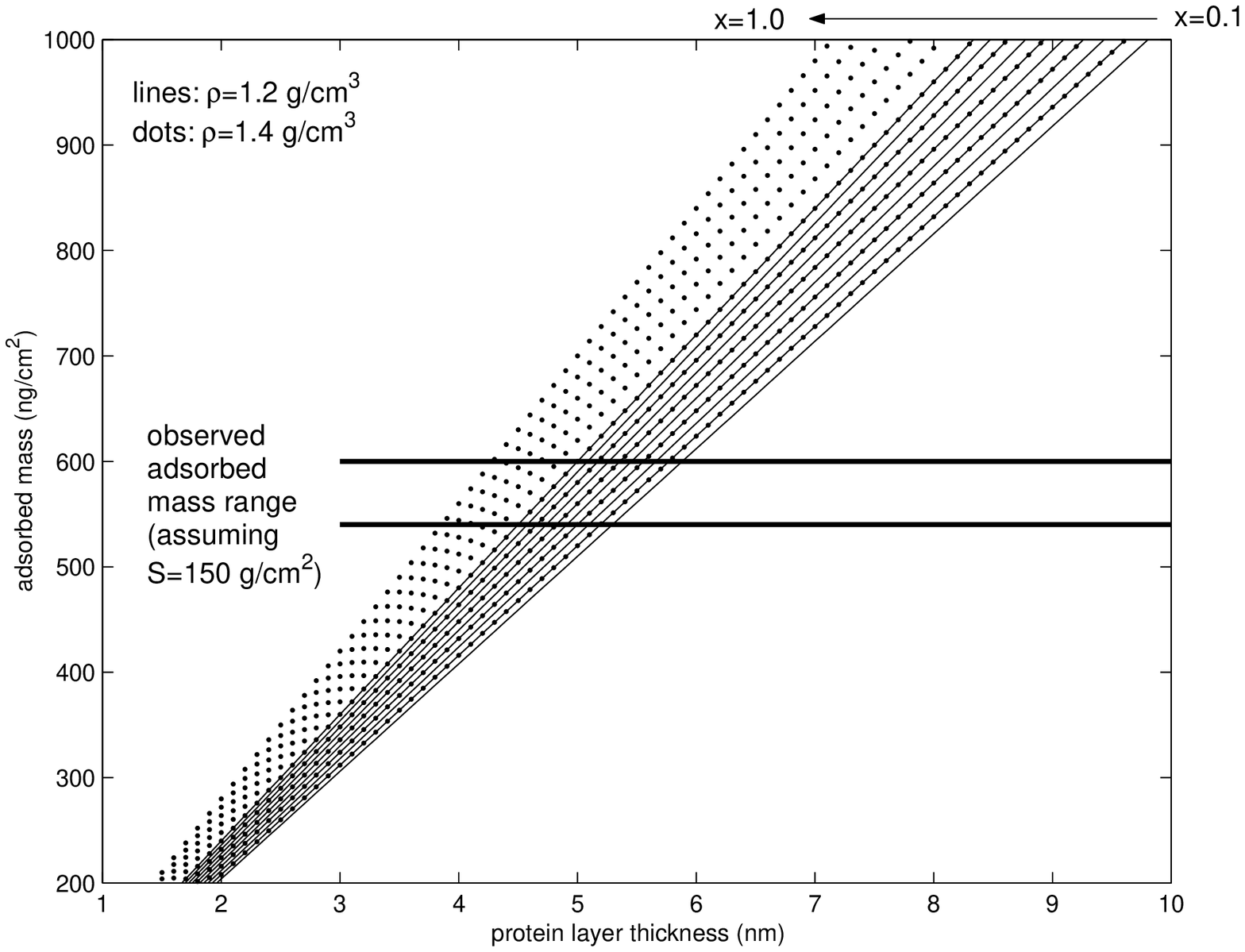} 
\includegraphics[width=15cm]{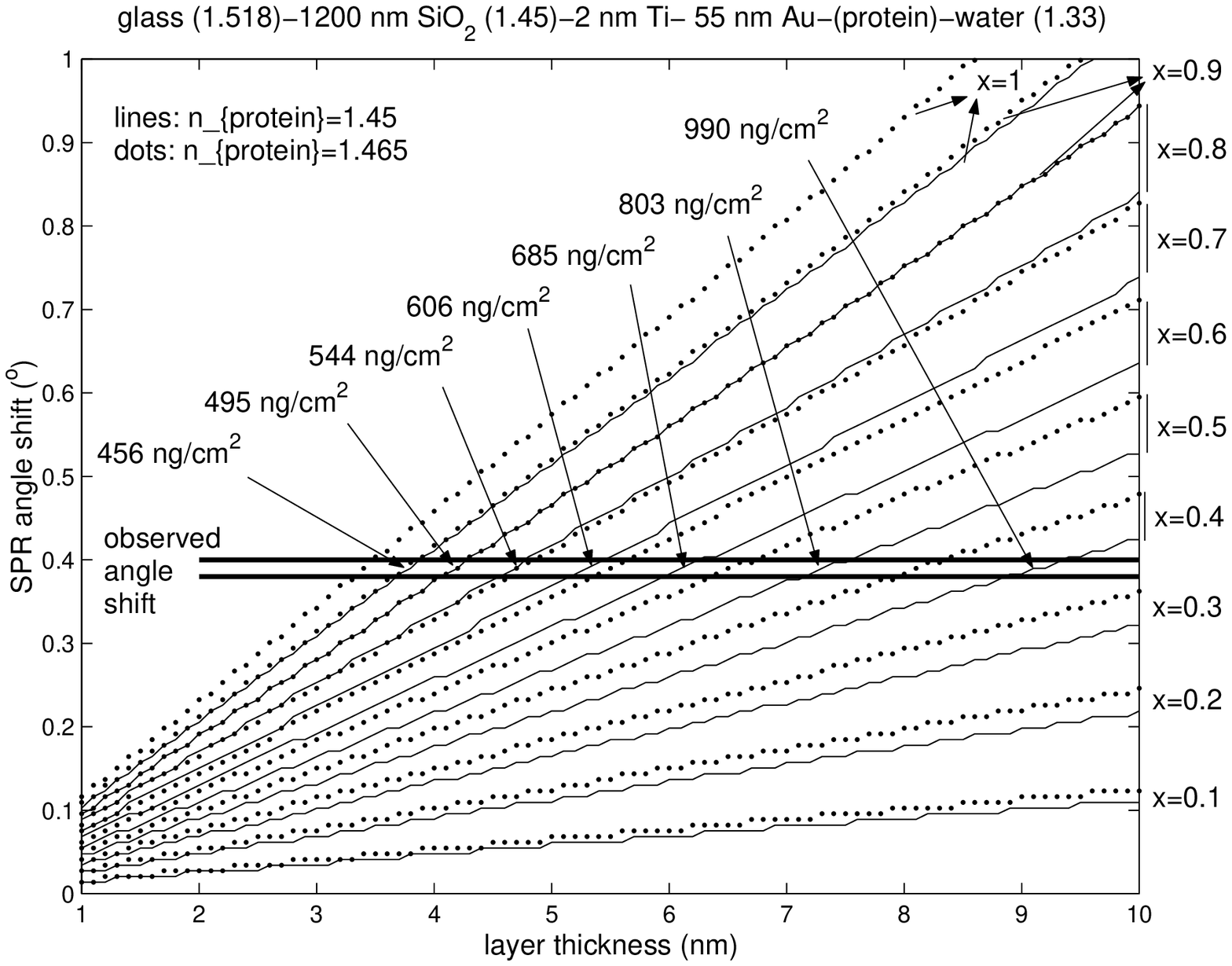}
\end{document}